\documentstyle[12pt,epsfig]{article}
\begin{document}
\begin{center}
{\Large \bf
Some new exact critical-point amplitudes
}

\bigskip
{\large H. Chamati$^{a)}$, D. M. Danchev$^{b)}$, N. S. Tonchev$^{a)}$}
\date{}
\smallskip

{\it $^{a)}$Georgy Nadjakov Institute of Solid State Physics - BAS,
Tzarigradsko chauss\'{e}e 72, 1784 Sofia, Bulgaria \\ $^{b)}$Institute
of Mechanics - BAS, Acad. G. Bonchev St. bl. 4, 1113 Sofia, Bulgaria }
\end{center}

\begin{abstract}
The scaling properties of the free energy and some of universal
amplitudes of a group of models belonging to the universality class of
the quantum nonlinear sigma model and the $O(n)$ quantum $\phi^4$
model in the limit $n\rightarrow \infty$ as well as the quantum
spherical model, with nearest-neighbor and long-range interactions
(decreasing at long distances $r$ as $1/r^{d+\sigma }$) is presented.
\end{abstract}

For temperature driven phase transitions quantum effects are
unimportant near critical points with $T_c>0$. However, if the systems
depends on another "non thermal critical parameter" $g$, at rather low
(as compared to characteristic excitations in the system)
temperatures, the leading $T$ dependence of all observables is
specified by the properties of the zero-temperature (or quantum)
critical point, say at $g_c$. The {\it dimensional crossover rule }
asserts that the critical singularities with respect to $g$ of a
$d$-dimensional quantum system at $T=0$ and around $g_c$ are {\it
formally} equivalent to those of a classical system with
dimensionality $d+z$ ($z$ is the dynamical critical exponent) and
critical temperature $ T_c>0$. This makes it possible to investigate
low-temperature effects (considering an effective system with $d$
infinite spatial and $z$ finite temporal dimensions) in the framework
of the theory of finite-size scaling. A compendium of some universal
quantities concerning $O(n)$-models at $n\to \infty$ in the context of
the finite-size scaling is presented.

\subsubsection*{Casimir amplitudes in critical quantum systems}
Let us consider a critical quantum system with a film geometry
$L\times\infty ^{d-1}\times L_\tau $, where $L_\tau=\hbar /(k_BT)$ is
the ``finite-size'' in the temporal (imaginary time) direction and let
us suppose that {\it periodic boundary conditions} are imposed across
the finite space dimensionality $L$ (in the remainder we will set
$\hbar =k_B=1$).

The confinement of {\it critical fluctuations} of an order parameter
field induces long-ranged force between the boundary of the plates
\cite{Fisher78,Krech94}. This is known as ``statistical-mechanical
Casimir force''. The Casimir force in statistical-mechanical systems
is characterized by the excess free energy due to the {\it finite-size
contributions} to the free energy of the bulk system. In the case it
is defined as
\begin{equation}
F_{{\rm Casimir}} (T,g,L|d)=-\frac{\partial f ^{{\rm ex}}(T,g,L|d)}{
\partial L},  \label{eq1}
\end{equation}
where $f ^{{\rm ex}}(T,g,L|d)$ is the excess free energy
\begin{equation}
f ^{{\rm ex}}(T,g,L|d)=f (T,g,L|d)-Lf(T,g, \infty|d). \label{eq2}
\end{equation}
Here $f(T,g,L|d)$ is the full free energy per unit area and per
$k_BT$, and $f(T,g,\infty|d)$ is the corresponding bulk free energy
density.

Then, near the quantum critical point $g_{c}$, where the phase
transition is governed by the non thermal parameter $g$, one could
state that ( see,~\cite{CDT99})
\begin{equation}
\frac1L f^{{\rm ex}}(T,g ,L|d)=\left( TL_\tau \right)
L^{-(d+z)}X_{{\rm ex}}^{{\rm u}}(x_1,\rho |d), \label{hypot1}
\end{equation}
with scaling variables
\begin{equation}
x_1=L^{1/\nu } \delta g,
 \ \mbox{and}\ \rho = L^z/L_\tau .
\end{equation}
Here $\nu$ is the usual critical exponent of the bulk model, $\delta g
\sim g-g_c$, and $ X_{{\rm ex}}^{{\rm u}}$ is the universal scaling
function of the excess free energy. According to the definition
(\ref{eq1}), one gets
\begin{equation}
F_{{\rm Casimir}}^{d}(T,g ,L)=\left( TL_\tau \right)
L^{-(d+z)}X_{{\rm Casimir}}^{{\rm u}}(x_1,\rho |d),
\label{defCasimir}
\end{equation}
where $X_{{\rm Casimir}}^{{\rm u}}(x_1,\rho |d)$ is the {\it
universal} scaling functions of the Casimir force.

It follows from Eq. (\ref{defCasimir}) that depending on the scaling
variable $\rho $ one can define Casimir amplitudes
\begin{equation}
\Delta _{{\rm Casimir}}^{{\rm u}}\left( \rho |d \right) := X_{{\rm
Casimir}}^{{\rm u}}\left( 0,\rho |d \right) . \label{DeltaDef}
\end{equation}

In addition to the ``usual'' excess free energy and Casimir
amplitudes, denoted by the superscript ``$u$'', one can define, in a
full analogy with what it has been done above, ``{\it temporal excess
free energy density''} $f_{{\rm t}}^{{\rm ex}}$,
\begin{equation}
f_{{\rm t}}^{{\rm
ex}}(T,g,|d)=f(T,g,\infty|d)-f(0,g,\infty|d) \label{deffext}.
\end{equation}
If the quantum parameter $g$ is in the vicinity of $g_c$, then one
expects
\begin{equation}
f_{{\rm t}}^{{\rm ex}}(T,g|d) =TL_\tau ^{- d/z}X_{{\rm ex}}^{{\rm
t}}\left( x_1^t|d\right), \label{defxextgen}
\end{equation}
i.e. instead of $X_{{\rm ex}}^{{\rm u}}(x_1,\rho |d)$. one has a
scaling function $X_{{\rm ex}}^{{\rm t}}\left( x_1^t|d\right)$ which
is the corresponding analog with scaling variables
\begin{equation}
x_1^t=L^{1/\nu z } \delta g .
\end{equation}
Obviously one can define the "temporal Casimir amplitude"
\begin{equation}
\Delta _{{\rm Casimir}}^{{\rm t}}\left( d\right)
:=
X_{{\rm ex}}^{{\rm t}}\left(0|d\right). \label{defCastgen}
\end{equation}

Whereas the ``usual'' amplitudes characterize the leading $L$
corrections of a finite size system to the bulk free energy density at
the critical point, the ``temporal amplitudes'' characterize the
leading temperature-dependent corrections to the ground state energy
of an {\it infinite} system at its quantum critical point $g_c$.

For the universality class under consideration the following exact
results are obtained:

(i)For the "usual" Casimir amplitudes
\begin{equation}
\Delta _{{\rm Casimir}}^{{\rm u}}\left( 0|2,2\right) =-\frac{2\zeta (3)}{
5\pi }\approx -0.1530, \label{eq14}
\end{equation}
here $\zeta (x)$ is the Riemann zeta function, and
\begin{equation}
\Delta^u_{\rm Casimir}(0|1,1)=-0.3157.
\label{deltasigma1}
\end{equation}

(ii)For the "temporal" Casimir amplitudes in the case ($0<\sigma \leq
2$)
\begin{equation}
\Delta _{{\rm Casimir}}^{{\rm t}}(\sigma ,\sigma )=-\frac{16}{5\sigma }
\frac{\zeta (3)}{(4\pi )^{\sigma /2}}\frac 1{\Gamma (\sigma /2)}.
\label{eq16}
\end{equation}

Note that the defined ''temporal Casimir amplitude'' $ \Delta
_{{\rm Casimir}}^{{\rm t}}(\sigma ,\sigma )$ reduces for $\sigma
=2$ to the ''normal'' Casimir amplitude $\Delta _{{\rm
Casimir}}^{{\rm u}}\left( 0|2,2\right) $, given by Eq.~(\ref{eq14}).
This reflects the existence of a special symmetry in that case between
the ''temporal'' and the space dimensionalities of the system.

When $\sigma\neq 2$ it is easy to verify that the following general
relation
\begin{equation}\label{dec}
\frac{\Delta _{{\rm Casimir}}^{{\rm t}}(\sigma ,\sigma )}{\Delta _{{\rm
Casimir}}^{{\rm t}}(2,2)}=\frac{8\pi}{\sigma(4\pi)^{\sigma/2}
\Gamma(\sigma/2)}
\end{equation}
between the temporal amplitudes holds. The r.h.s. of (\ref{dec}) is a
decreasing function of $\sigma$.

\paragraph*{Relation with the Zamolochikov's $C$-function}

Let us note that if $z=1$ the temporal excess free energy introduced
above coincides, up to a (negative) normalization factor, with the
proposed by Neto and Fradkin definition of the non-zero temperature
generalization of the $C$-function of Zamolodchikov (see e.g. Ref.
\cite{DT99}).

For $z\ne 1$ a straightforward generalization of this definition can
be proposed at least in the case of long-range power-low decaying
interaction
\begin{equation}
C(T,g|d,z)=-T^{-(1+d/z)} \frac {v^{d/z}}{n(d,z)} f_{{\rm ex}}^{{\rm
t}}(T,g|d),
\label{Cgen}
\end{equation}
where $z=\sigma/2$, $v=TL_{\tau}$ and
\begin{equation}
n^t(d,\sigma )=\frac 4\sigma \frac{\zeta \left(1+2d/\sigma\right)}{(4\pi
)^{d/2}}\frac {\Gamma (2d/\sigma)} {\Gamma(d/2)}.
\label{nt}
\end{equation}

\begin{figure}
\centerline{\epsfig{file=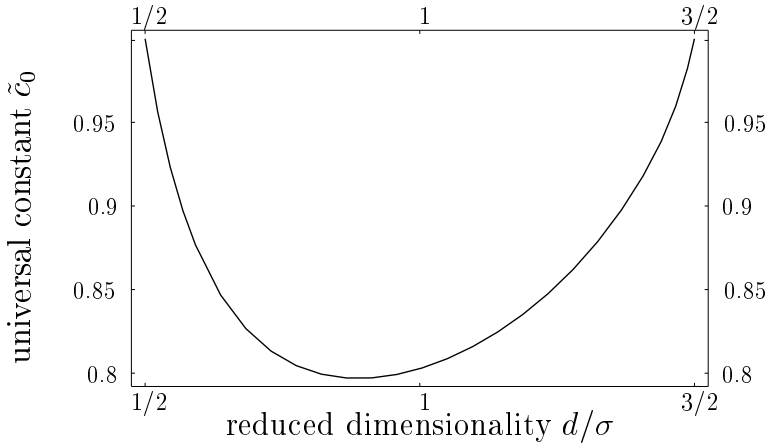,width=4in}}
\caption{Behaviour of the universal constant $\tilde c$ as
a function of $\frac{d}{\sigma}$.}
\label{fig1}
\end{figure}

The quantity ${\tilde c}_0(d,\sigma ):=C(T,g_c|d,z)$ is an important
universal characteristic of the theory. The behavior of ${\tilde
c}_0(d,\sigma )$ is calculated numerically for dimensions between the
lower critical dimension $\sigma/2$ and upper critical dimension
$3\sigma/2$ for arbitrary values of $0<\sigma\leq2$. The results are
universal as function of $d/\sigma$ as it is presented on
Fig.\ref{fig1}. In the particular case $d/\sigma=1$, one can obtain
analytically \cite{CDT99}
\begin{equation}
\tilde c_0(\sigma,\sigma) =4/5.
\label{ff}
\end{equation}
This generalizes the result obtained for $d=\sigma=2$ \cite{Sachdev93}
to the case of long-range interaction.

\begin{figure}
\centerline{\epsfig{file=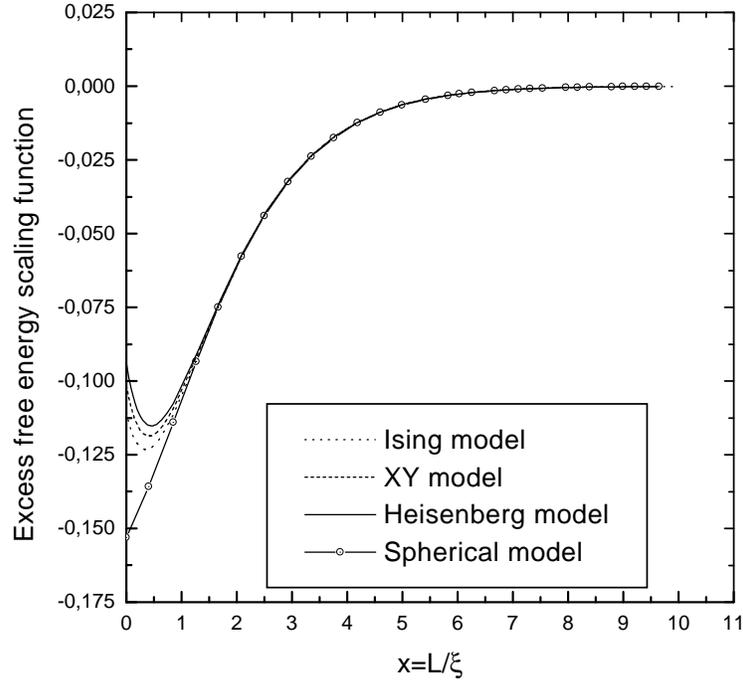,width=3.83in}}
\caption{The {\it universal} zero-field finite-size scaling
functions $X^{ex}$ of the excess free energy as a function
of the scaling variable $x=L/\xi(T>T_c)$ for Ising, XY,
Heisenberg and Spherical models. }
\label{comparison}
\end{figure}

To shed some light to what extend the amplitudes presented above are
close to that one of more realistic models we present a comparison of
the scaling functions of the excess free energy of the Ising, XY,
Heisenberg and spherical model (limit $n\to\infty$) in
FIG.~\ref{comparison}. The results for the spherical model are exact
while that ones for the Ising, XY and Heisenberg models are obtained
by $\epsilon $-expansion technique up to the first order in $\epsilon
$. The Monte Carlo results for he 3d Ising model give $-0.1526\pm
0.0010$ \cite{K97}, which is {\it surprisingly close} to the exact
value~(\ref{eq14}). This makes difficult to resolve the question how
$X^{ex}/n$ approaches the corresponding result for the spherical model
when $n\rightarrow
\infty $. Note that all the curves practically overlap for $L>2\xi $,
where $\xi$ is the correlation length.

\subsubsection*{Other amplitudes}

Other important universal critical amplitudes, in finite-size scaling,
depend upon the geometry $L_{d-d'}\times\infty^{d'}\times L_\tau^z$ as
well as the range of the interaction. One of the most important
quantities for a numerical analysis is the Binder's cumulant ratio.
For the quantum 2d spherical model with $\sigma=2$ at the critical
point it is \cite{D98}
\begin{equation}
B=\frac{2\pi }{\sqrt{5}\ln ^3\tau }\approx 25.21657,  \label{bv}
\end{equation}
where $\tau=(1+\sqrt{5})/2$ is the "golden mean" value.

In what follows we will list a number of results obtained in the
framework of the quantum spherical model~\cite{CPT98} and the ${\cal
O}(n)$ quantum $\varphi^4$ model \cite{CT99}.

\noindent ({\bf i}) Finite system at zero temperature:

\begin{equation}
d=\sigma=1: \ \ \ \ \ \ \ \ \ \ \frac{L}{\xi}=0.624798 \ \ \ \ \ {\rm for} \ \ \ d'=0.
\end{equation}

\begin{equation}
d=\sigma=2: \ \ \ \ \ \ \ \ \ \  \frac{L}{\xi}=\left\{
\begin{array}{ll}
1.511955 & {\rm for} \ \ \ \ d^{\prime}=0, \\
0.962424 & {\rm for} \ \
\ \ d^{\prime}=1.
\end{array}
\right.
\end{equation}

\begin{figure}
\centerline{\epsfig{file=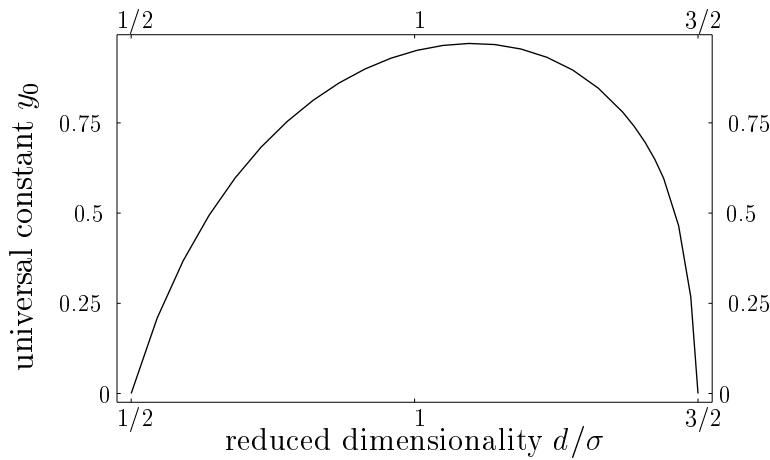,width=4in}}
\caption{Behaviour of the scaling variable $y_0=L_\tau/\xi$
at the quantum critical point as a function of $\frac{d}{\sigma}$.}
\label{fig3}
\end{figure}

\noindent ({\bf ii}) Bulk system at finite temperature:

\begin{equation}
d=\sigma: \ \ \ \ \ \ \ \ \ \ \ \ \ \ \ \frac{L_\tau}{\xi}= 0.962424.
\end{equation}
This result is a just a point in graph presented in FIG.~\ref{fig3},
where we show the behaviour of $L_\tau/\xi$ as a universal function of
the ratio $d/\sigma$. The point corresponding to $(\frac
d\sigma=1,y_0=0.962424)$ can be obtained analytically~\cite{CT99}.

The above result are obtained for the case when the quantum parameter
controlling the phase transition is fixed at its critical value. Now
we will present results obtained when the quantum parameter is fixed
by ``running'' values corresponding the shifted critical quantum
parameter. We are limited to the case $d=\sigma=2$
\begin{equation}
\frac{L}{\xi}=\left\{
\begin{array}{ll}
7.061132 & {\rm for} \ \ \ \ d^{\prime}=1, \\
4.317795 & {\rm for} \ \
\ \ d^{\prime}=0,
\end{array}
\right.
\end{equation}
for finite system at zero temperature and
\begin{equation}
\frac{L_\tau}{\xi}=\left\{
\begin{array}{ll}
7.061132 & {\rm for} \ \ \ \ d^{\prime}=1, \\
6.028966 & {\rm for} \ \
\ \ d^{\prime}=0,
\end{array}
\right.
\end{equation}
for the bulk system at finite temperature~\cite{CPT98}.

\bigskip
\bigskip
This work is supported by The Bulgarian Science Foundation (Projects
F608/96 and MM603/96).

\end{document}